# Unveiling the GeI$_2$-Assisted Oriented Growth of Perovskite Crystallite for High-Performance Flexible Sn Perovskite Solar Cells

Huagui Lai,[a] Selina Olthof,[b] Shengqiang Ren,[c,*] Radha K. Kothandaraman,[a] Matthias Diethelm,[d] Quentin Jeangros,[e] Roland Hany,[d] Ayodhya N. Tiwari,[a] Dewei Zhao,[c] and Fan Fu [a,*]

a. Laboratory for Thin Films and Photovoltaics, Empa – Swiss Federal Laboratories for Materials Science and Technology, Ueberlandstrasse 129, CH-8600 Duebendorf, Switzerland.

b. Department of Chemistry, University of Cologne, Greinstrasse 4–6, 50939, Cologne, Germany.

c. College of Materials Science and Engineering, & Engineering Research Center of Alternative Energy Materials & Devices, Ministry of Education, Sichuan University, Chengdu 610065, Sichuan, China.

d. Laboratory for Functional Polymers, Empa – Swiss Federal Laboratories for Materials Science and Technology, Ueberlandstrasse 129, CH-8600 Duebendorf, Switzerland.

e. Photovoltaics and Thin Film Electronics Laboratory, Institute of Electrical and Microengineering, École Polytechnique Fédérale de Lausanne, CH-2002 Neuchâtel, Switzerland

* Corresponding author: fan.fu@empa.ch, rensq@scu.edu.cn

## Abstract

Tin perovskites are emerging as promising alternatives to their lead-based counterparts for high-performance and flexible perovskite solar cells (PSCs). However, their rapid crystallization often leads to inadequate film quality and poor device performance. In this study, the role of GeI$_2$ as an additive is investigated for controlling the nucleation and crystallization processes of formamidium tin triiodide (FASnI$_3$). The findings reveal the preferential formation of a Ge-rich layer at the bottom of the perovskite film upon the introduction of GeI$_2$. It is proposed that the initial formation of the Ge-complex acts as a crystallization regulator, promoting oriented growth of subsequent FASnI$_3$ crystals and enhancing overall crystallinity. Through the incorporation of an optimal amount of GeI$_2$, flexible Sn PSCs with an efficiency of 10.8% were achieved. Furthermore, it was observed that the GeI$_2$ additive ensures a remarkable shelf-life for the devices, with the rigid cells retaining 91% of their initial performance after more than 13,800 hours of storage in an N$_2$ gas environment. This study elucidates the mechanistic role of GeI$_2$ in regulating the nucleation and crystallization process of tin perovskites, providing valuable insights into the significance of additive engineering for the development of high-performance flexible tin PSCs.

**Introduction**

Halide perovskites have emerged as highly promising materials for thin film photovoltaics (PVs), owing to a combination of remarkable merits such as tunable direct bandgaps, excellent optoelectronic properties, high defect tolerance, etc.[1] Lead-based perovskites, in particular, have been extensively studied, with record power conversion efficiencies (PCEs) of 26.1%.[2] However, toxic lead in perovskites raises environmental and health concerns that may hinder their widespread application, especially in flexible and portable electronics. As a less toxic alternative, tin (Sn)-based perovskites have recently gained more attention. These compounds degrade into non-toxic compounds after exposure to ambient air. They can be safely applied to niche applications, such as wearable electronics, building-integrated PVs, and indoor PVs.[3] However, the research and development of Sn-based perovskite solar cells (PSCs) is more challenging than that of their Pb-based counterparts. The main problems are the fast crystallization kinetics of Sn perovskites and easy oxidation of $Sn^{2+}$ to thermodynamically more stable $Sn^{4+}$, which can hinder the formation of high-quality and uniform perovskite film.[4] To cope with these challenges, many strategies, such as composition,[5] additive,[6] and solvent engineering,[7] etc., have been extensively carried out with Sn-based systems. Among the varied strategies, additive engineering has been reported to be a practical approach for controlling the fast crystallization of the Sn-based perovskites and suppressing the easy oxidation of $Sn^{2+}$. Various chemicals, such as $SnF_2$,[8] 1,4-bis(trimethylsilyl)-2,3,5,6-tetramethyl-1,4-dihydropyrazine (TM-DHP),[9] ethylenediammonium diiodide ($EDAI_2$)[10] and phenylhydrazine hydrochloride (PHCl),[8b] etc., have been used as additives in Sn-based perovskite precursor, either to work as a reducing reagent or as a crystallization regulator to cope with the issues mentioned above.

Germanium (II) iodide ($GeI_2$) has been used as an effective additive, first reported by Nozomi Ito et al. to get a mixed Sn-Ge perovskite, and an enhanced device performance was obtained.[11] It was found that the small ionic radii of Ge allows it to passivate the top surface of the Sn perovskite. In a recent study, an amorphous $GeO_2$ layer was found to form at the interface of $NiO_x/FASn_{0.9}Ge_{0.1}I_3$ and was helpful to passivate the bottom interface and further improve device performance.[11-12] However, the distribution of $GeI_2$ in $FASnI_3$ and its potential to form Sn-Ge perovskites, as opposed to its possible preference for localization at the surface or bottom of perovskite films, still presents an unresolved and ambiguous question.

Here in this work, we first elucidate the effect of GeI$_2$ on the growth of Sn perovskite film, where the proper amount of GeI$_2$ leads to improved crystallinity and, therefore, better optoelectronic properties. We observe that there is no X-ray diffraction (XRD) peak shifting for the Sn perovskites upon GeI$_2$ introduction, which indicates no Sn-Ge perovskite forming and GeI$_2$ is therefore acting more like an impurity in Sn perovskites. We investigate the surface chemistry of the Sn perovskites with X-ray photoelectron spectroscopy (XPS), which confirms a suppressed neutral iodide content and oxidation of surface Sn$^{2+}$ in the presence of GeI$_2$ additive. Additionally, we have precisely determined the distribution of Ge within the Sn perovskite film for the first time by utilizing time-of-flight secondary ion mass spectroscopy (ToF-SIMS). Surprisingly, we find a preferential accumulation of Ge at the bottom of perovskite film with a low concentration of GeI$_2$ and at the top of the perovskite surface when increasing the concentration of GeI$_2$. Combined with XPS and TOF-SIMS results, we confirm a suppressed F$^-$ distribution at the perovskite surface, which indicates a strong interaction between the GeI$_2$ and SnF$_2$ additives. Based on our findings, we propose a bottom-up crystallization model to explain particular distribution of Ge. Our model suggests earlier nucleation and crystallization of Ge-based complexes during the film formation, serving as a seed layer for the growth of highly oriented FASnI$_3$ perovskite. This enables a more controlled crystallization process and ultimately improves crystal quality. By incorporating appropriate quantities of GeI$_2$ additives, we have successfully fabricated flexible Sn-based PSCs with a record PCEs of 10.8%. In addition, the Sn-based PSCs with GeI$_2$ additives also exhibited excellent shelf lifetime, maintaining over 91% of their initial performance after 13,800 hours of storage in N$_2$ condition.

**Results and discussion**

We first study the effect of GeI$_2$ additives in Sn perovskites on glass substrates for ease of processing. GeI$_2$ was added in varying concentrations (0%, 1%, 2.5%, 5%, 7.5%, and 10% molar ratio to SnI$_2$) to the FASnI$_3$ perovskite precursors. Additionally, 10% SnF$_2$ (molar ratio to SnI$_2$) was added to all precursors as a common additive to prevent Sn$^{2+}$ oxidation. Upon preparation, we observed that the solutions became increasingly milky with the addition of GeI$_2$ (**Figure S1**), likely attributed to the limited solubility of GeI$_2$ in DMF and DMSO solvents.

**Structural, morphological, and optical properties**

**Figure 1a** presents top-view scanning electron microscopy (SEM) images of Sn-perovskites grown on glass/ITO/PEDOT:PSS substrates. With increased GeI$_2$ additives, distinct changes in film morphologies can be observed. The domain size distribution, depicted in **Figure 1b**, reveals that the average grain size tends to be maximized with 5% GeI$_2$ additives (Mean value of domain size see **Figure S2**). However, in the 7.5% and 10% groups, the averaged domain sizes are smaller, accompanied by white precipitation, likely indicating the formation of secondary phases. The atomic force microscopy (AFM) results align with the SEM observations, confirming the influence of GeI$_2$ on the domain size (**Figure S3**). A similar impact was also observed for Sn-perovskite films grown on PEN/ITO/PEDOT: PSS substrates (**Figure S4**).

**Figure 2a** shows the X-ray diffraction (XRD) patterns for the perovskite films with different amounts of GeI$_2$ additives (The full patterns are shown in **Figure S5**). The XRD peak at around 14° corresponds to the (100) planes of the orthorhombic structure of the FASnI$_3$ perovskite.[10, 13] Intensities and full width at half maximum (FWHM) of this peak were extracted and plotted in **Figure 2b**. With 5% GeI$_2$ additive, the peak intensity of the (100) plane was increased by more than two-fold compared to the 0% GeI$_2$ group, while two-fold decreased in the FWHM. These results prove that the GeI$_2$ additive can enhance the crystallinity of the FASnI$_3$ perovskite. Importantly, no apparent peak shifting was observed in the XRD patterns of the FASnI$_3$ perovskites with increasing amounts of GeI$_2$ additives. This result confirms that Ge has not been incorporated into the Sn perovskite lattice, which differs from the previous report.[11] Moreover, a diffraction peak at around 12.7° was observed for the film with a 10% GeI$_2$ additive. We performed XRD

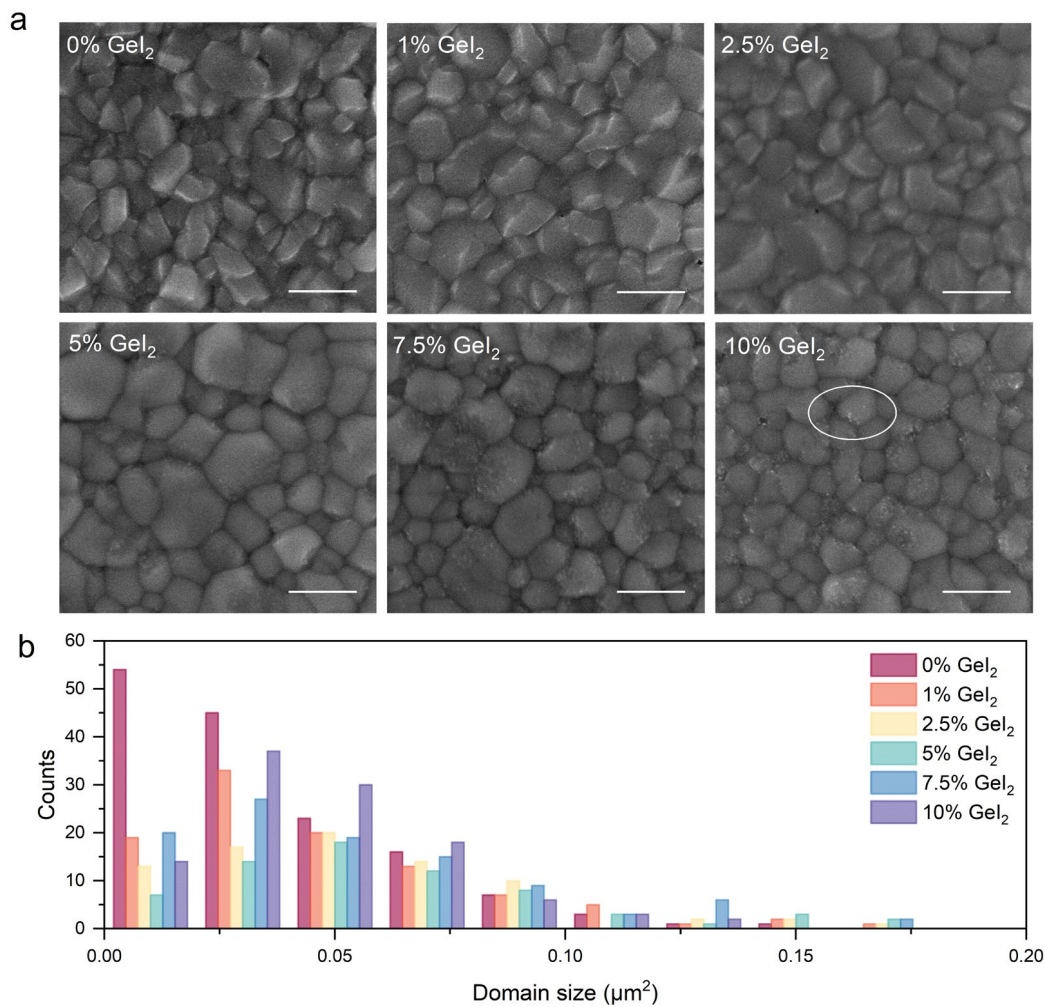

Figure 1 (a) Top-view scanning electron microscopy (SEM) images for Sn perovskite films with different concentration of GeI$_2$. Scale bar, 500 nm. White precipitation is circled out. (b) Domain size distribution extracted from SEM images.

characterization for pure SnI$_2$, GeI$_2$, and SnF$_2$ films to confirm its origin, as shown in **Figure S6**. The SnI$_2$ film shows identical XRD peaks at around 12.7° and 25.7° as that of the additional XRD peaks observed for the perovskite film with 10% GeI$_2$ additive, as shown in **Figure 2a**. It is evident that with excess GeI$_2$ additives, the SnI$_2$ phase is forced to form at the perovskite surface.

We checked the absorbance of the perovskite films by measuring their reflectance and transmittance, as shown in **Figure 2c**. By adding the GeI$_2$ additive, the absorbance of the perovskite films in the 500–900 nm range is close, while the 0% and 5% GeI$_2$ group perovskite films show higher values in the 350-500 nm range. We performed time-resolved photoluminescence (TrPL) characterization on the perovskite films to gain insights into the carrier lifetime. **Figure 2d** shows the TrPL decay curves fitted with a single exponential function. We found that the film with 5% GeI$_2$ additive exhibited a significantly improved PL lifetime τ of 12.83 ns compared to the reference film (τ=4.15 ns). Markedly increased PL lifetime suggests reduced non-radiative recombination in the Sn-perovskite film. However, it is worth noting that the films with 7.5% and 10% GeI$_2$ showed a slightly decreased PL lifetime. This could be attributed to the negative effects caused by the excess SnI$_2$ on the film surface, which counteracts the positive impact brought by GeI$_2$.

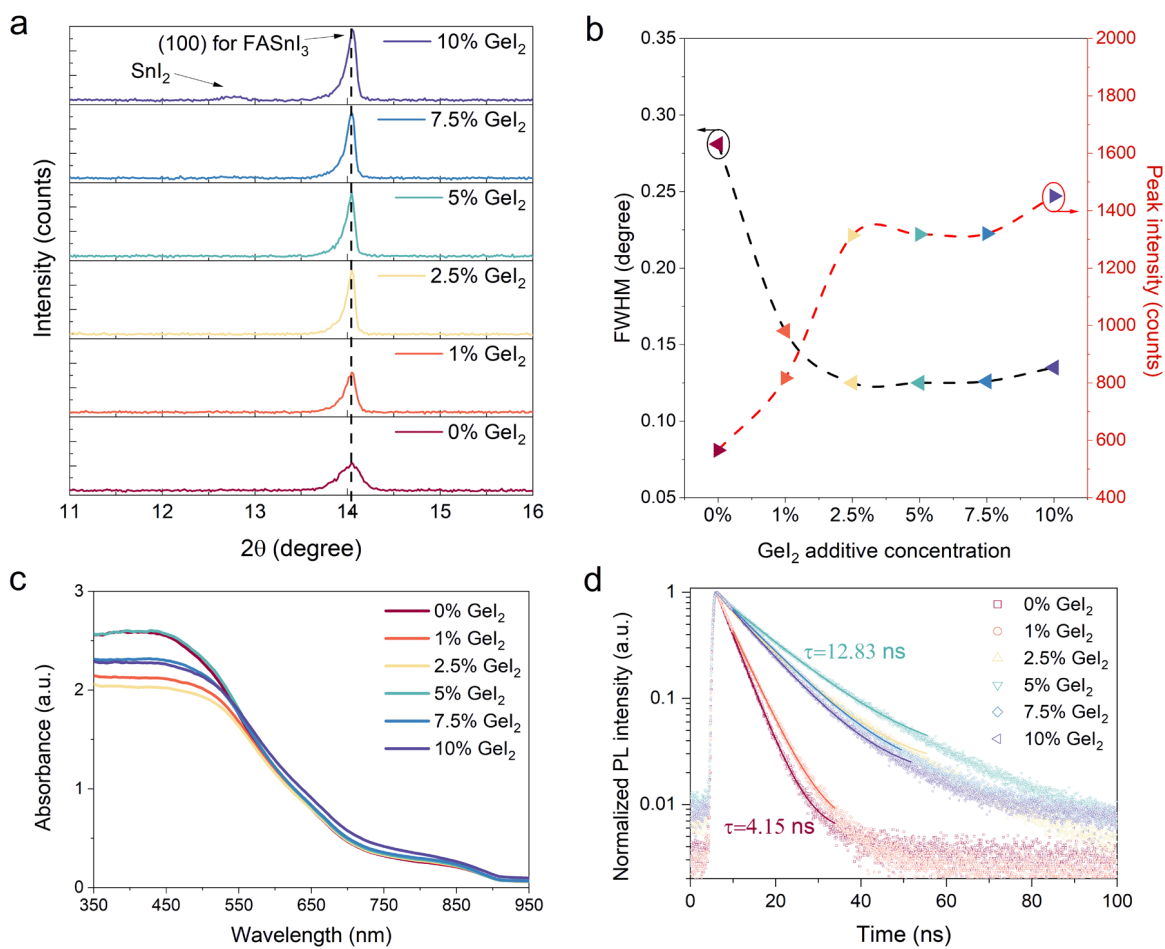

Figure 2 (a) X-ray diffraction (XRD) patterns for the Sn perovskite films with different concentrations of GeI$_2$ and corresponding (b) XRD intensity and full width at half maximum of the (100) peak. (c) Absorbance spectra and (d) Time-resolved photoluminescence decays for the Sn perovskite films with different concentrations of GeI$_2$.

**Chemical state and elements distribution**

We proceeded to investigate the impact of GeI$_2$ on the oxidation of FASnI$_3$ perovskites using surface sensitive XPS analysis. In **Figure 3a**, analysis of the XPS core level peak of iodide revealed the presence of a small amount of more neutral iodide on the surface of the reference perovskite film. Interestingly, the addition of GeI$_2$ resulted in a decreased amount of neutral iodide, suggesting that GeI$_2$ either acts as a preventive measure against the oxidation of I$^-$ or facilitates the reduction of neutral I$_2$ to I$^-$. This effect is advantageous as it reduces vacancies, thereby enhancing device performance and long-term stability.[14] Fitting the Sn feature in XPS is challenging as Sn can appear in various bonds and oxidation states in these films. In **Figure 3b**, we have followed our recently established fitting procedure for SnF$_2$-containing tin perovskite films.[15] The fit includes three features that are assigned to Sn$^{2+}$ in FASnI$_3$ perovskite, degraded Sn$^{4+}$, as well as newly formed fluorinated SnO$_2$ species. Notably, although all films employed SnF$_2$ as an anti-oxidant, the reference perovskite film is the only sample exhibiting a peak associated with Sn$^{4+}$. In contrast, all the films with GeI$_2$ additive showed no Sn$^{4+}$ peak. This finding indicates that using SnF$_2$ alone does not entirely suppress the oxidation of Sn$^{2+}$, while the additional inclusion of GeI$_2$ effectively mitigates the formation of Sn$^{4+}$. These results highlight the crucial role of the GeI$_2$ additive in FASnI$_3$ as an anti-oxidant, reducing the levels of neutral iodide and Sn$^{4+}$, which are considered detrimental to the perovskite film.

Interestingly, we also observed a significant decrease in the intensity of the fluorine (F$^-$) signal detected on the perovskite surface with increasing concentrations of GeI$_2$ additive, as depicted in **Figure 3c**.

To validate this observation, ToF-SIMS was conducted to examine the distribution of elements within the perovskite films. Consistent with the findings from XPS data, adding GeI$_2$ resulted in a significant reduction in the F$^-$ signal detected on the surface of the perovskite films, as shown in **Figure 3d**. To quantitatively compare the impact of the GeI$_2$ additive on the change in F$^-$ signal intensity at the perovskite surface, we extracted the corresponding F$^-$ signal values and presented them as a bar chart in **Figure 3d**. Our analysis revealed that adding just 1% GeI$_2$ already resulted in a substantial 70% reduction in F$^-$ intensity compared to the reference sample with 0% GeI$_2$. Moreover, subsequent increments in GeI$_2$ concentration led to only minor variations in F- intensity, demonstrating that the reduction in F$^-$ signal intensity saturates beyond a specific GeI$_2$ concentration.

In **Figure 3e**, we furthermore evaluate the Ge distribution within the perovskite film. For perovskite film with 1% of GeI$_2$ additive, the result indicates predominant Ge accumulation at the bottom of the perovskite film, with a minimal presence at the upper surface and the bulk perovskite. Upon increasing the GeI$_2$ additive to 5%, we note a modest Ge accumulation at the film's top surface, while no Ge is detected within the bulk of perovskite. With increasing concentration of GeI$_2$, the distribution of Ge shifts in the layer. As we can see from the ToF-SIMS result for the perovskite film with 10% GeI$_2$, Ge is still mostly found at the bottom and top layer, however, a small amount is detectable throughout the bulk of perovskite. Specifically, we reveal that Ge most readily precipitates at the bottom interface, and with increasing amount it first starts appearing on the surface and followed by emergence throughout the bulk of the perovskite.

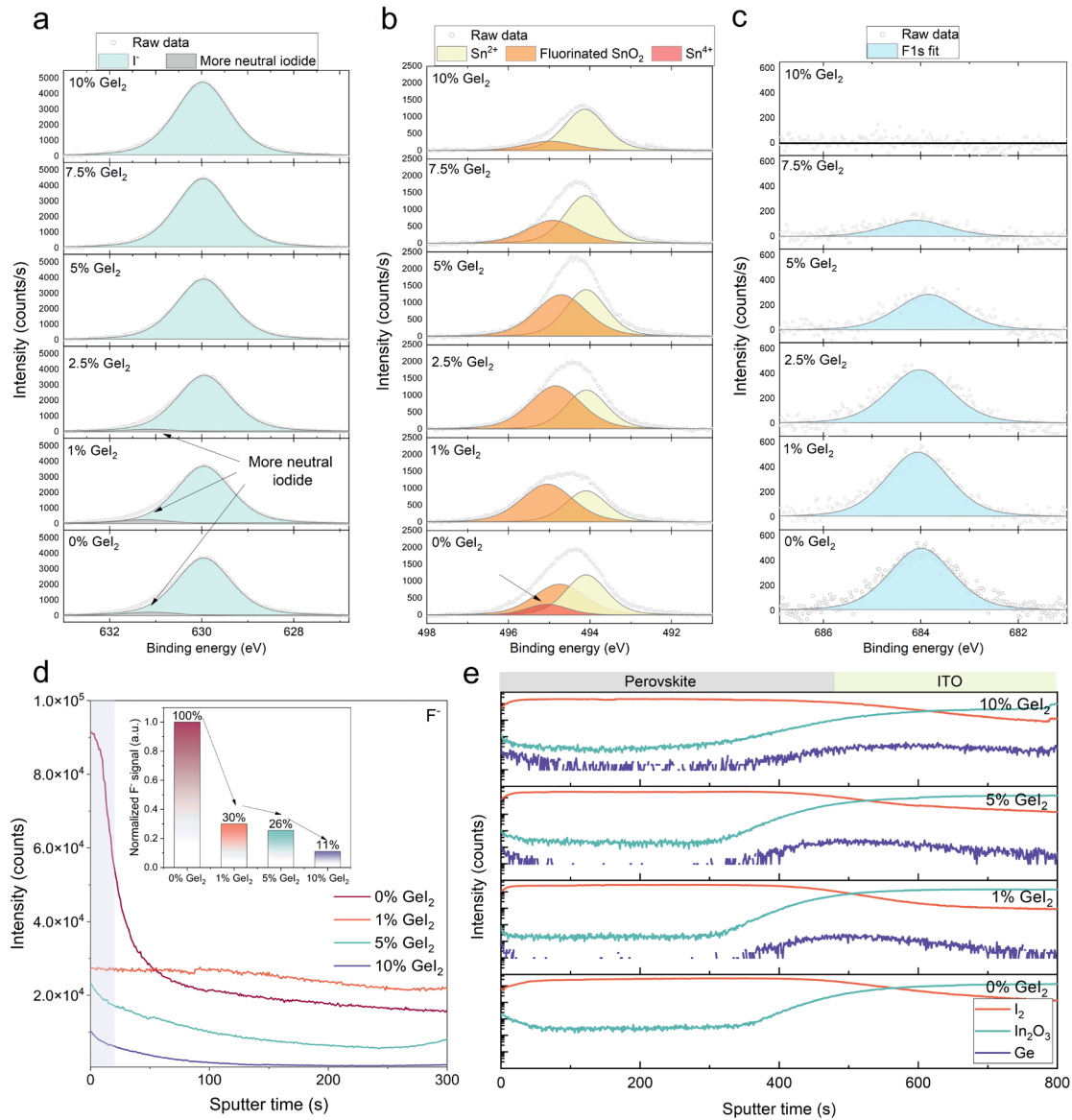

Figure 3 XPS measurements of the surface of Sn-based perovskites with different concentrations of GeI$_2$, showing the core level peaks of (a) I 3d$_{3/2}$, (b) Sn 3d$_{3/2}$, and (c) F 1s, respectively. (d) Depth profiling of F$^-$ throughout the Sn perovskite films, with a bar chart presenting normalized F$^-$ signal (using 0% GeI$_2$ group as reference) at the perovskite surface. (e) Depth profiling of Ge, I$_2$, and In$_2$O$_3$ throughout the Sn perovskite films on ITO.

**Formation of Ge distribution**

To understand the unique distribution of Ge in the perovskite film, we have proposed a bottom-up crystallization model that considers the formation of Ge-complexes with varying solubility.

As shown in **Figure 4**, a bottom-to-up crystallization model is proposed to elucidate the nucleation and crystallization from liquid phase to solid phase during the spin-coating and annealing of perovskite precursors with GeI$_2$ additive. We didn't observe any peak shifting based on XRD patterns (**Figure 2a**), indicating that Ge has not been incorporated into the FASnI$_3$ perovskite crystals. For impurities incorporated outside the crystals, the mechanism is mostly attributed to the generation of a supersaturated state that drives the impurity precipitation.[16] Combined with the finding of less F$^-$ detected at the perovskite surface with GeI$_2$ additive, we hypothesize that the GeI$_2$ may form a low-soluble complex with SnF$_2$ in the liquid system before nucleation. During the nucleation, when the spin-coating and anti-solvent dripping occur, the solvent is rapidly extracted, forming a supersaturated state for the low-soluble Ge-complexes. Due to their limited solubility, these complexes enter the supersaturated stage earlier than other components, resulting in a significant accumulation of the Ge-complexes at the bottom of the perovskite film. We propose that the precipitated Ge-complexes act as seed crystals that preferentially interact with the (100) face of the FASnI$_3$ perovskites. This preference may arise from differences in growth rate or surface chemistry, such as exposed functional groups. The presence of these seed crystals contributes to the high quality of the perovskite film during crystallization.[16]

When the concentration of the GeI$_2$ additive is increased, high-soluble Ge-complexes can remain in the liquid phase due to a limited supply of SnF$_2$, as inferred from ToF-SIMS results (**Figure 3d**). Consequently, these high-soluble Ge-complexes are repulsed away from the solid-liquid interface during the crystallization and tend to accumulate at the top surface due to the higher growth rate of FASnI$_3$. This may explain the emergence of a Ge signal on the top surface of the perovskite film.

However, it is important to note that at high concentrations of GeI$_2$ (e.g., 10% GeI$_2$ additive), an excess accumulation of Ge-complexes at the bottom can lead to the formation of excessive nucleation sites, resulting in competitive crystal growth and, consequently, smaller average grain size. This bottom-up crystallization model underscores the critical importance of selecting appropriate additives to effectively control the type and quantity of seed crystals. This control, in turn, enables the precise growth of perovskite crystals characterized by oriented phases, larger grain sizes, and superior quality.

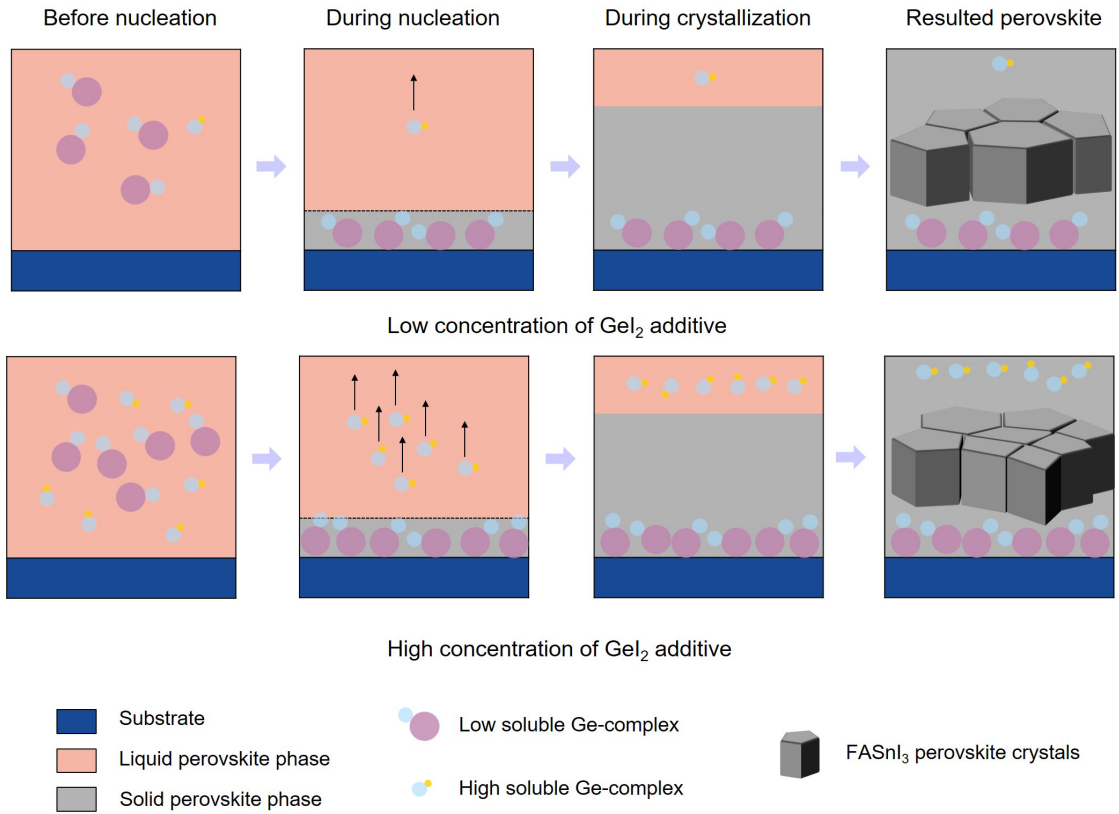

Figure 4 Schematics show the formation of special distribution of Ge during the spin-coating and annealing of perovskites.

**Charge dynamics and deivce performance**

To evaluate the PV performance of the Sn perovskites within solar cells, we fabricated flexible solar cells in a p-i-n architecuture based on Sn perovskite films with 0% and 5% GeI$_2$ additives. The photo of the flexible Sn PSCs is shown in **Figure 5a**. We performed the cross-sectional high-angle annular dark-field scanning transmission electron microscopy (HAADF-STEM) imaging as well as the energy-dispersive X-ray spectroscopy (EDX) mapping for the device with 5% GeI$_2$ as shown in **Figure 5b**. From the EDX mapping, we can see that the Ge was detected in the PEDOT:PSS layer, which supports the conclusion that Ge most readily precipitates at the bottom interface and therefore becoming incorporated into the PEDOT:PSS layer. We performed impedance spectroscopy (IS) to gain insight into the charge transportation process of different PSCs. **Figure 5c** shows impedance analysis employing an equivalent circuit model. The Nyquist plots show only a single semicircle, representing the devices' charge recombination resistance ($R_{rec}$).[17] The 5% GeI$_2$ device offers a much larger $R_{rec}$ (~90k Ohm) than the 0% GeI$_2$ device (~25k Ohm). This larger $R_{rec}$ for the 5% GeI$_2$ device originates from a more oriented crystal and fewer defect-assisted traps, indicating a suppressed charge recombination. The charge carrier lifetime was further analyzed via transient photovoltage (TPV) measurements performed under open-circuit conditions, as shown in

**Figure 5d**. The decay times of the photovoltage were fitted with a single exponential decay model. A much longer decay time for the 5% GeI$_2$ device (96.7 us) than that for the 0% GeI$_2$ device (46.7 us) can be explained by the suppressed nonradiative recombination and the better film quality brought by the GeI$_2$.

We further conducted current density-voltage (*J-V*) measurements on the solar cells and evaluated the PCE of multiple devices. The summarized PCE distribution is shown in **Figure S7**. The devices incorporating 5% GeI$_2$ additive exhibited an average PCE of 8.5%, much higher than the 6.0% achieved by the 0% GeI$_2$ group. It is worth noting that our devices exhibited a light-soaking behavior, resulting in improved PV performance after multiple *J-V* measurements. This phenomenon, previously observed in Sn PSCs, is attributed to strain relaxation in the perovskite material.[18] **Figure 5e** shows the *J-V* curves of the flexible Sn PSCs after light-soaking. The champion device based on 5% GeI$_2$ additive achieves an enhanced PCE of 10.8% under forward scan, with 0.716 V in open-circuit voltage ($V_{OC}$), 20.76 mA cm$^{-2}$ in short-circuit current density ($J_{SC}$) and 72.9% in fill factor (FF). In contrast, the device with 0% GeI$_2$ shows a PCE of 8.2% under forward scan, with 0.560 V of $V_{OC}$, 20.14 mA cm$^{-2}$ of $J_{SC}$, and 72.81% of FF. The improved $V_{OC}$ indicates reduced non-radiative recombination within the perovskite film when the GeI$_2$ additive is present. The steady power output of the solar cells was measured as shown in **Figure S8**, with PCEs of ~7.2% and ~10.72% for the 0% and 5% GeI$_2$ solar cells, respectively. Furthermore, we performed external quantum efficiency (EQE) measurements for the 0% and 5% GeI$_2$ devices, as depicted in **Figure 5f**. The integrated $J_{SC}$ values obtained from the EQE spectra are 19.13 and 21.00 mA cm$^{-2}$ for the 0% and 5% GeI$_2$ devices, respectively, consistent with the $J_{SC}$ values obtained from the *J-V* curves. Overall, we have achieved an efficiency of 10.8% in flexible, lead-free Sn PSCs by incorporating 5% GeI$_2$ additive. This achievement marks the highest recorded efficiency within this particular category, as highlighted in **Figure 5g.**

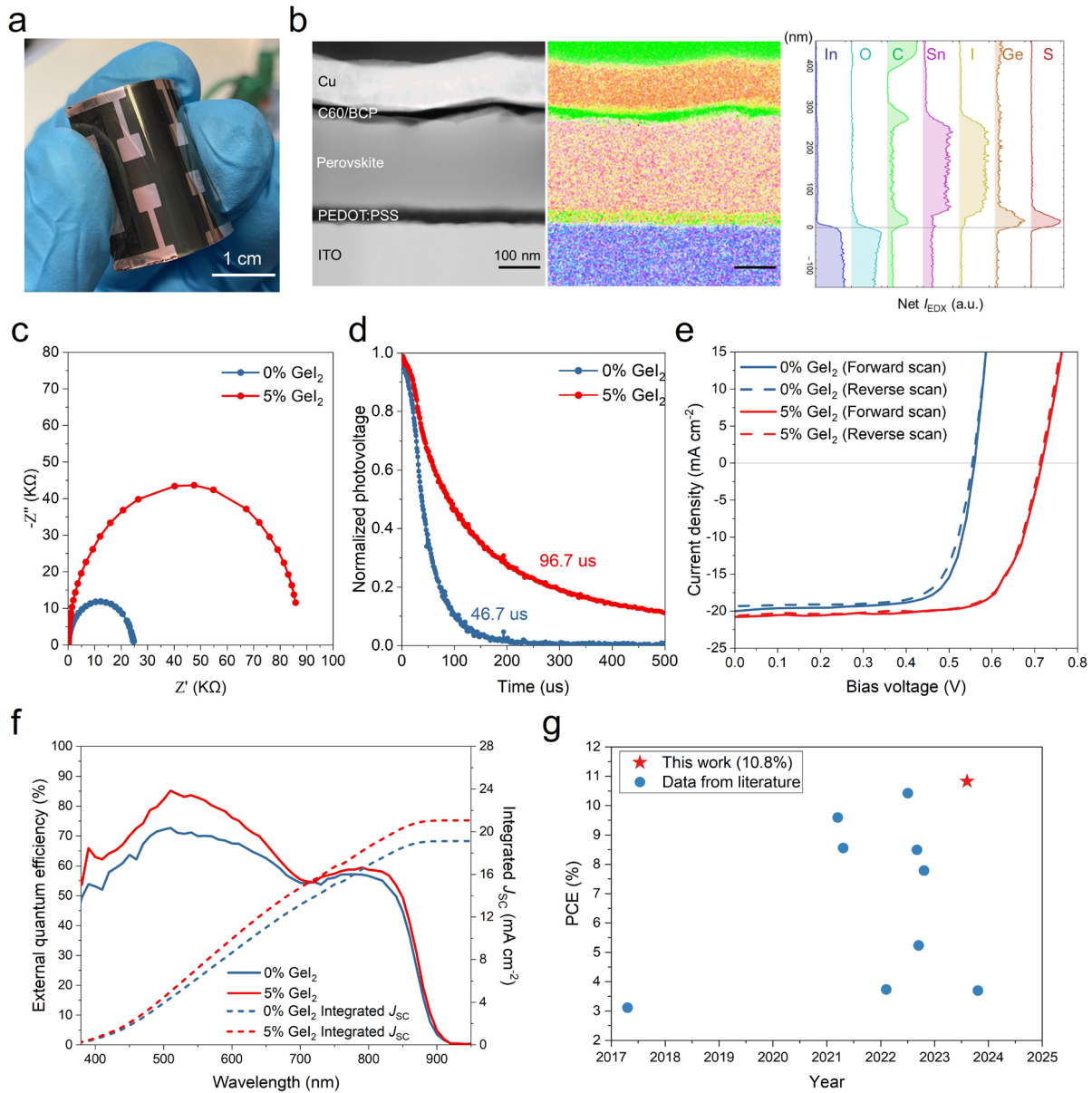

Figure 5 (a) Photo of the Sn perovskite solar cells on a flexible substrate. (b) Cross-sectional HAADF STEM imaging and the corresponding energy-dispersive X-ray (EDX) mapping and line profiles for the device with 5% GeI$_2$. (c) Impedance spectroscopy (IS), (d) Transient photovoltage (TPV) measurements for the devices with 0% and 5% GeI$_2$, respectively. (e) J-V curves and (f) EQE spectra for the devices with 0% and 5% GeI$_2$, respectively. (g) Efficiency progress of flexible Sn-based perovskite solar cells.[12, 19]

We also conducted a light-soaking stability study on the encapsulated solar cells, as shown in **Figure 6a**. To rule out the impact of flexible substrates, we used rigid Sn PSCs to perform the experiment. By employing continuous maximum power point tracking (MPPT), we observed that the 5% GeI$_2$ additive device could maintain approximately 80% of its initial efficiency after about 150 hours light soaking (~42 °C). Furthermore, we investigated the long-term shelf-life stability of multiple rigid devices containing 5% GeI$_2$ additive. These devices, on average, maintained over 91% of their initial performance when stored in an nitrogen (N$_2$) glove box over a period of ~13,800 hours (574 days) (**Figure S9**). **Figure 6b** presents typical J-V curves of the rigid Sn PSCs with 5%

GeI$_2$ during long-term shelf lifetime tracking. Compared to the fresh state, we observed an initial increase in the $V_{OC}$ of the device from 0.757 V to 0.826 V, leading to a peak PCE of 13.48% after 7 days of storage. After 574 days of tracking, the primary degradation was observed in the FF, which decreased from 78.43% to 72.22%. Overall, this device maintained over 95% of its initial PV performance after 574 days, representing a significant advancement in developing stable Sn PSCs. These findings highlight the potential of GeI$_2$ as an effective additive for enhancing the Sn-based PSCs' performance and long-term stability.

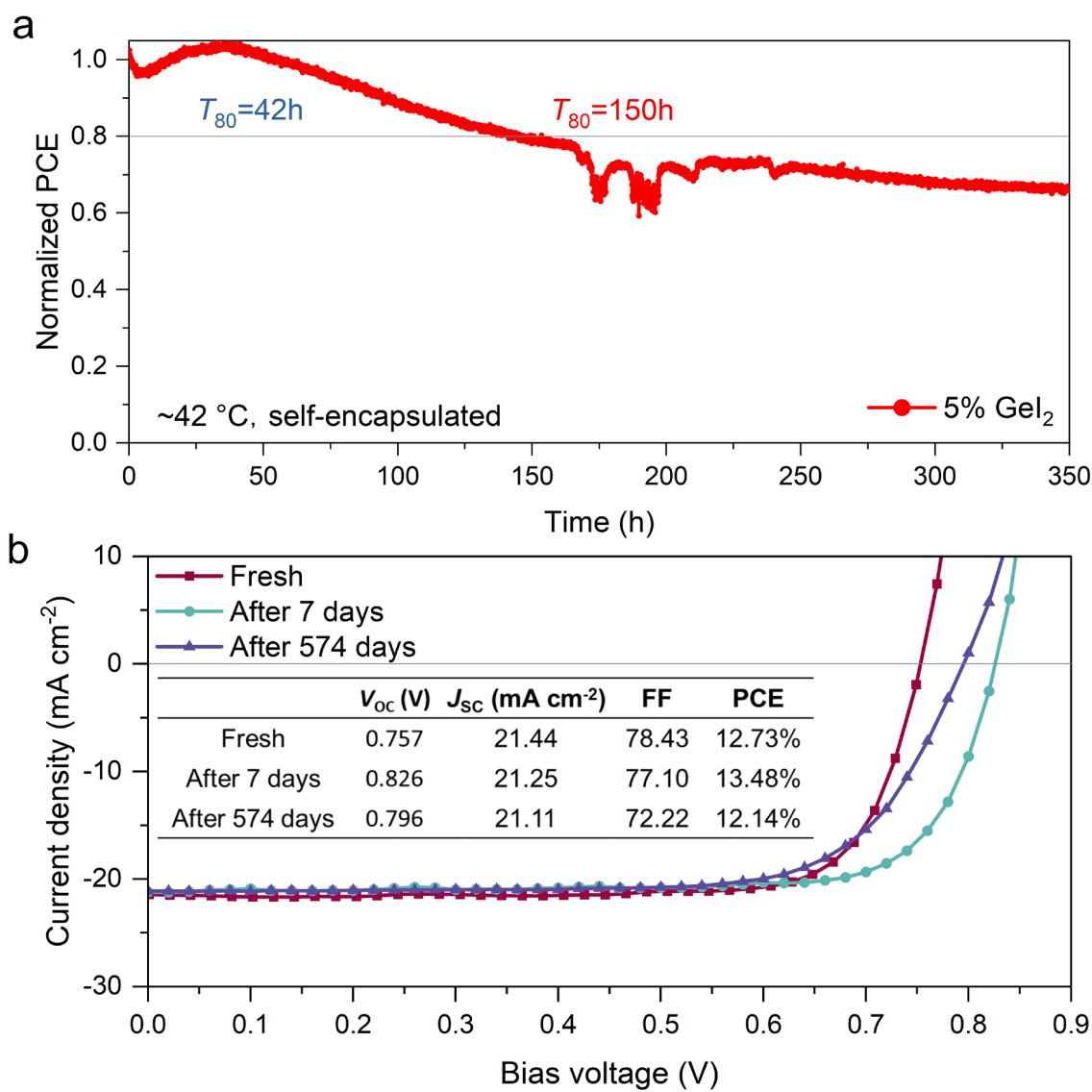

Figure 6 (a) Maximum power point (MPP) tracking of the rigid Sn perovskite solar cells with 5% GeI$_2$. (b) Typical *J-V* curves for rigid Sn perovskite solar cells with 5% GeI$_2$ during shelf lifetime tracking, *J-V* parameters inserted.

## Conclusions

Our study investigated the role of GeI$_2$ as an additive in FASnI$_3$ perovskite films. The impact of GeI$_2$ on the structure and morphology of the perovskite films was examined, revealing that the incorporation of GeI$_2$ facilitates the crystallization process, leading to preferred crystal growth and high crystal quality. Interestingly, we show a sequential precipitation pattern, with Ge initially concentrating at the bottom, followed by emergence at the top surface, culminating in widespread distribution within the bulk of the perovskite. Based on these observations, we propose a bottom-up crystallization model to explain the distribution of Ge, highlighting the aggregated Ge layer as a seed layer that influences and controls the crystallization process. By incorporating GeI$_2$ into the Sn-based perovskite, we achieved a record PCE of 10.8% for flexible Sn-based PSCs. Remarkably, the GeI$_2$-added PSCs also demonstrated impressive long-term stability. Our findings contribute to understanding the effects of additives in perovskite precursors and provide valuable insights for accelerating the development of flexible Sn PSCs.

## Author Contributions

F.F. and H.L. conceived the idea. H.L. fabricated the films/devices and performed various characterizations. S.O. conducted the XPS measurement. R.K.K conducted the XRD and AFM measurement. S.R. performed the TrPL measurements. M.D. and R.H. supported with device characterizations. Q.J. performed STEM measurements. H.L. and F.F. wrote the manuscript with input from all co-authors. A.N.T., S.R., D.Z., and F.F directed and supervised the overall project. All authors discussed the results and reviewed the manuscript.

## Conflicts of interest

There are no conflicts to declare.

## Acknowledgements

This work has received funding from the European Union's Horizon Europe research and innovation programme under grant agreement No 101075605 (SuPerTandem). This work has also been financially supported by Swiss National Science Foundation (grant no. 200021_213073) and Swiss Federal Office of Energy (SFOE, grant no. SI/502549-01). This work is partially financially supported by the National Natural Science Foundation of China (nos. 62104163 and 62174112) as well as the Natural Science Foundation of Sichuan Province (no. 2022NSFSC1183). Huagui Lai thanks the China Scholarship Council (CSC) funding from the Ministry of Education of P. R. China.